\documentclass[12pt]{article}

\usepackage{scicite}

\usepackage{times}

\usepackage{amsmath}
\usepackage{amsfonts}
\usepackage{amssymb}
\usepackage{graphicx}

\newenvironment{sciabstract}{%
\begin{quote} \bf}
{\end{quote}}

\title{Successive cohorts of Twitter users \\show increasing activity and shrinking content horizons}

\author
{Frederik Wolf,$^{1,2\ast\dagger}$ Philipp Lorenz-Spreen,$^{3\ast\dagger}$ Sune Lehmann$^{4,5\dagger}$\\
\\
\normalsize{$^{1}$Potsdam Institute for Climate Impact Research (PIK)}\\
\normalsize{- Member of the Leibnitz Association, Telegrafenberg, 14473 Potsdam, Germany}\\
\normalsize{$^{2}$Department of Physics, Newtonstraße 15, Humboldt University, 12489 Berlin, Germany}\\
\\
\normalsize{$^{3}$Center for Adaptive Rationality,}\\
\normalsize{Max Planck Institute for Human Development, Berlin, 14195, Germany}\\
\\
\normalsize{${^4}$Department of Applied Mathematics and Computer Science,}\\
\normalsize{Technical University of Denmark, Lyngby, DK-2800, Denmark}\\
\\
\normalsize{$^{5}$Center for Social Data Science, University of Copenhagen, Copenhagen, 1353, Denmark}\\
\\
\normalsize{$^\ast$These two authors contributed equally}\\
\\
\normalsize{$^\dagger $ E-mails: Frederik Wolf -- frederik.wolf@pik-potsdam.de,}\\
\normalsize{Philipp Lorenz-Spreen -- lorenz-spreen@mpib-berlin.mpg.de, Sune Lehmann -- sljo@dtu.dk}
}

\date{}

\begin{document} 


\baselineskip24pt


\maketitle



\begin{sciabstract}
The global public sphere has changed dramatically over the past decades: 
A significant part of public discourse now takes place on algorithmically driven platforms owned by a handful of private companies.
Despite its growing importance, there is scant large-scale academic research on the long-term evolution of user behaviour on these platforms, because the data are often proprietary to the platforms.
Here, we evaluate the individual behaviour of 600,000 Twitter users between 2012 and 2019 and find empirical evidence for an acceleration of the way Twitter is used on an individual level. This manifests itself in the fact that cohorts of Twitter users behave differently depending on when they joined the platform.
Behaviour within a cohort is relatively consistent over time and characterised by strong internal interactions, but over time behaviour from cohort to cohort shifts towards increased activity. Specifically, we measure this in terms of more tweets per user over time, denser interactions with others via retweets, and shorter content horizons, expressed as an individual's decaying autocorrelation of topics over time. 
Our observations are explained by a growing proportion of active users who not only tweet more actively but also elicit more retweets. 
These behaviours suggest a collective contribution to an increased flow of information through each cohort's news feed---an increase that potentially depletes available collective attention over time.
Our findings complement recent, empirical work on social acceleration, which has been largely agnostic about individual user activity. 
\end{sciabstract}

\section*{Introduction}
Year by year, the world is becoming more interconnected online \cite{Data19}, with news \cite{vosoughi2018spread}, games \cite{morcos2019internet}, and entertainment~\cite{ribeiro2020auditing} delivered to individuals via an increasing number of smartphones and computers worldwide \cite{taylor2019smartphone}.
The realization that this development may not be unequivocally beneficial for individuals and societies has spurred an active scientific and public debate \cite{zuboff2019age, aral2020hype, hills2019dark, bail2018exposure, allen2020evaluating, allcott2020welfare, mosleh2021shared}, while positive consequences of the growing connectivity can also be observed \cite{guess2019less, barbera2015tweeting, yang2020exposure, Boxell10612}. 
Still, most aspects of the complex interplay between information technology, social interconnectedness, and human behaviour on the collective\cite{bak2021stewardship}, as well as on the individual level \cite{Lorenz-Spreen20} have yet to be empirically addressed through large-scale quantitative studies. 

One crucial and overarching concept, which is discussed particularly in sociology, is a development termed ``social acceleration''\cite{rosa2013beschleunigung}. Social acceleration is described as the interplay between the dimensions of technological acceleration, acceleration of social change and the acceleration of the pace of life\cite{rosa2003social}. Our findings complement recent, empirical work on social acceleration, which did not address individual user activity. 

The dimension of technological acceleration has been quantified in a variety of sectors, from genomic sequencing\cite{wetterstrand2013dna} to computing power\cite{moore1998cramming} and transmission of information\cite{hilbert2011world}. 
The other dimensions concerning the impact of such technological developments on the social sphere are more difficult to quantify empirically\cite{wajcman2008life}. 
But gradually more and more empirical hints for the presence of social acceleration are emerging. Recent work has provided evidence for acceleration of collective attention across various domains, including information search, communication, and entertainment \cite{lorenz2019accelerating}. These findings are supported by other, empirical evidence for instances of social acceleration, for example, on media consumption and production \cite{hutchins2011acceleration}, the editing style of Hollywood movies\cite{cutting2011quicker}, the take-up of new concepts in books \cite{michel2011quantitative}, even the uptake of technological innovations itself\cite{mcgrath2013pace} and, most related, the information consumption on social media \cite{yang2020exposure, Scharkow2020, ford2021competition}. 

The majority of existing work, however, focuses on the aggregated, collective level, leaving open the question of whether there is a general accelerating trend on the individual (with a few exceptions, like surveys that point to a reduction in sustained attention in reading behaviour \cite{liu2005reading}). 

In other words, are the observed developments driven simply by the fact that there are more people participating in consumption or discourse---and that they became more visible there? Or do people behave differently now compared with just a few years ago? Here we aim to address this question by illuminating and quantifying one aspect of social acceleration, namely whether and how individuals use Twitter differently over time.



We addressed these questions from the perspective of changing behaviour on Twitter, using a longitudinal data set from which we sampled $600,000$ randomly chosen individual users with a total of $83,706,899$ retweet interactions, within and outside that random sample. 
This data enabled us to measure user behaviour over 8 years (2012--2019), spanning a large fraction of the observable period of widespread social media adoption. 
Based on this data set, we analyzed individual activity on Twitter from 2012 to 2019, aiming to understand differences between users who joined the platform at different points in time.
We also explored other aspects of individual behaviour ---namely, the development of interactions with others and the amount of time users pay attention to topics. 

Our work connects to recent efforts to understand the long-term changes of (and on) social media that have only now become quantitatively accessible, despite the widespread use of social media over the last decade \cite{Waller2020, Alshaabi2020}. 
 
\begin{figure*}
\centering
\includegraphics[width=17.8 cm]{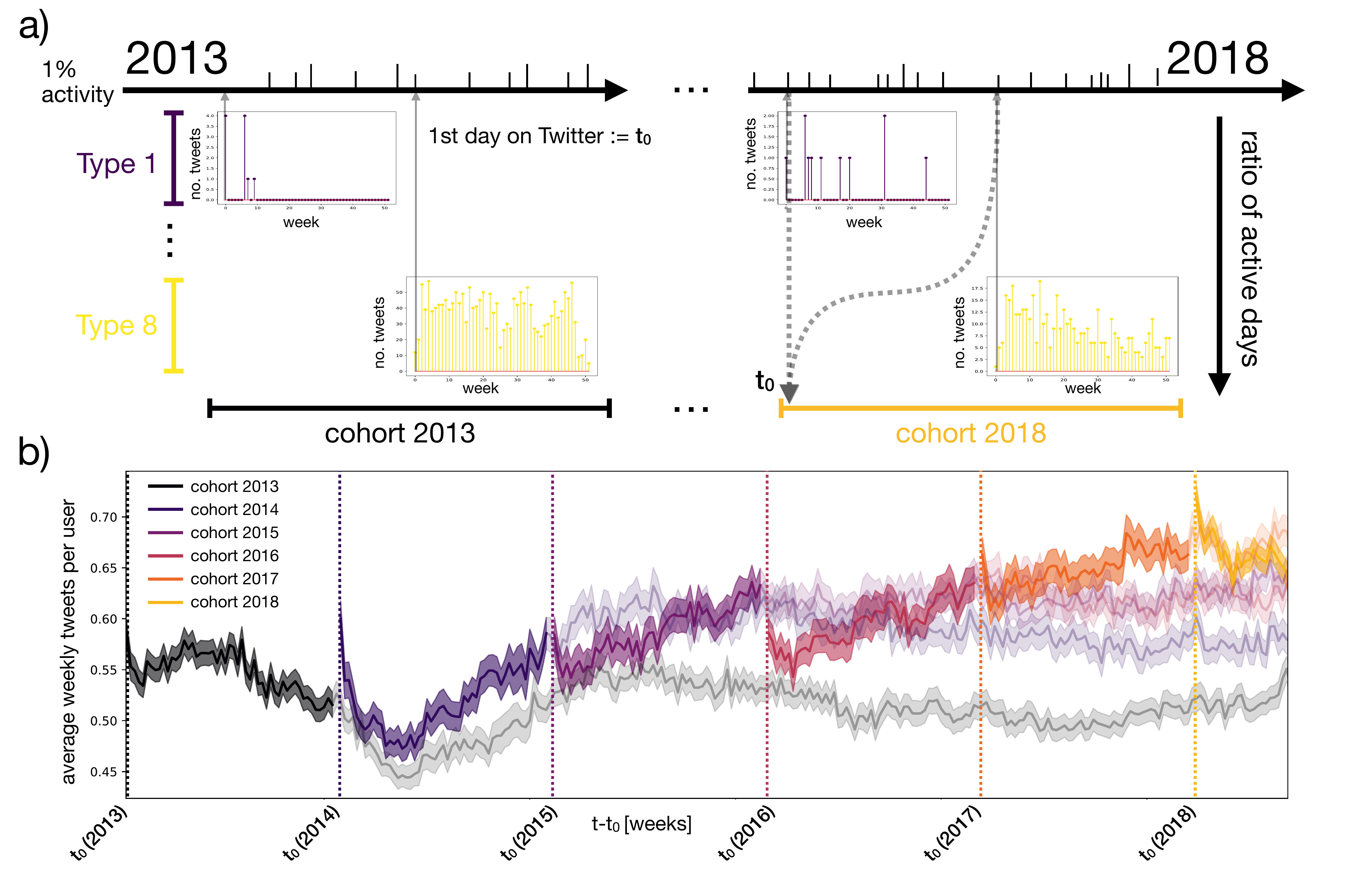}
\caption{User cohorts and user types in the Twitter data set. a) Illustration of $1\%$ downsampling from the original tweet trajectory of a single user between $2013$ and $2019$, which was necessary due to the reduced availability through Twitter's Decahose API. Illustration of the difference between user cohorts (horizontal axes) and user types (vertical axes). User types are sorted by their individual ratio of active to total days on Twitter. b) Average tweets per week and user within each cohort, where the day of the first tweet is aligned per cohort, showing an average Twitter career starting at $t_0$ and lasting until 2019 for all users, due to our sampling strategy. First years highlighted for comparison, subsequent years shaded.}
\label{fig:1}
\end{figure*}

\section*{Results}
A longitudinal dataset from Twitter's Decahose API allowed us to pursue this question (see Material and Methods section for details about the dataset). We sample user who have been active at least twice, once in a specific month 2019 and once in the past after 2013 (see also the Material and Methods section for details on the sampling strategy). In combination with personal identifiers, this ensures that active Twitter users can be tracked over the entire period and that the observed effects are not driven by people who had stopped using the platform. To start, we analyze the activity of users in their first active year (year after first recorded tweet or retweet) on Twitter. For a first insight into the general trend we run a simple linear regression between the individual starting date and the individual mean inter-event time (time between own tweets or retweets of the first, active year). We found first indications of individual acceleration: the inter-event time has decreased by approximately $6.5$h per year between $2013$ and $2018$ (considering the available $1\%$ sample). 

For more detailed analyses, the heterogeneity of individual behaviour over time make it necessary to pool users in order to create meaningful average observational data at the systemic level.
We choose two strategies, dividing users into groups by the year they began using Twitter and by user type (Fig.~\ref{fig:1}a).
Forming groups of users based on the year they were first active on Twitter allows to divide them into eight \textit{cohorts} that began using Twitter in the different years (2012,2013,...2019).
Dividing users into groups of comparable user types according to their activity level makes it possible to look beyond the simple cohort average and characterize the changing composition of behaviours across cohorts.

\subsection*{Users grouped by cohort}

When dividing users into cohorts based on the year they were first active on Twitter, we find that the tweeting behaviour of people who started using Twitter in 2013 is substantially different from that of people who started actively using Twitter more recently.

To assess the change in user activity over time, we first identify the date $t_0$ of each user's first recorded tweet, then split the users into cohorts according to their starting year. 
To compare the behaviour of users who had spent the same amount of time on Twitter, we align all users from each cohort by setting their starting dates $t_0$ to an arbitrary but common date. 
This makes it possible to approximately represent individual behaviour as an average Twitter experience within a cohort and to evaluate possible differences between cohorts. 
Fig.~\ref{fig:1}a illustrates this procedure.

We then calculate the mean number of tweets per week for all users at the same stage of Twitter use, for each week of the observation period. 
Aligning individual users creates an offset of the trajectories within each cohort, which could be up to one year. 
To avoid this offset to extend beyond the active period of a user from our sample, we ignore the year 2019 for each cohort, leaving at least one year of buffer (and did not consider the 2019 cohort). 
Additionally, as we had no data from before 2012 and thus could not know whether users had joined Twitter before our observation period, we ignore the cohort from 2012. We interpret inactivity throughout 2012 as a proxy for not having joined Twitter earlier.

Fig.~\ref{fig:1}b shows the resulting average tweeting activity per user for all cohorts from 2013 to 2018---thereby going beyond merely illustrating the growth in the number of users. This visualization highlights both the dissimilarity between cohorts at the same stage of Twitter use and an offset in user activity (measured as average tweets per week) after multiple cohorts. There is a clear trend of increasing activity on the platform from one cohort to the next, while activity levels remain stable over long periods within each cohort of users (Fig.~\ref{fig:1}b). 

While user behaviour changes between cohorts, it remains relatively stable within cohorts. Users who were active on Twitter in 2013 are still using the platform in a way that is similar to when they started. They also connect preferentially to users who join around the same time. More generally, an average of 90\% of all retweets occur within a single cohort, despite increasing total interactions, indicating homophily among contacts of the same cohort. This finding mirrors previous findings of politically homophilous ties in social media networks \cite{Moslehe2022761118} and the formation of topical groups therein \cite{Cinellie2023301118}. 

Users' initial activity increases in each subsequent cohort and remains stable (and increasing) at a higher level of activity, especially for the cohorts after 2015. Furthermore, the cohorts end up on significantly different levels of activity at the end of our observation period.

What drives this social acceleration, and what other dimensions of behaviour changed? 

\subsection*{Users grouped by activity}

\begin{figure*}
\centering
\includegraphics[width=13.4 cm]{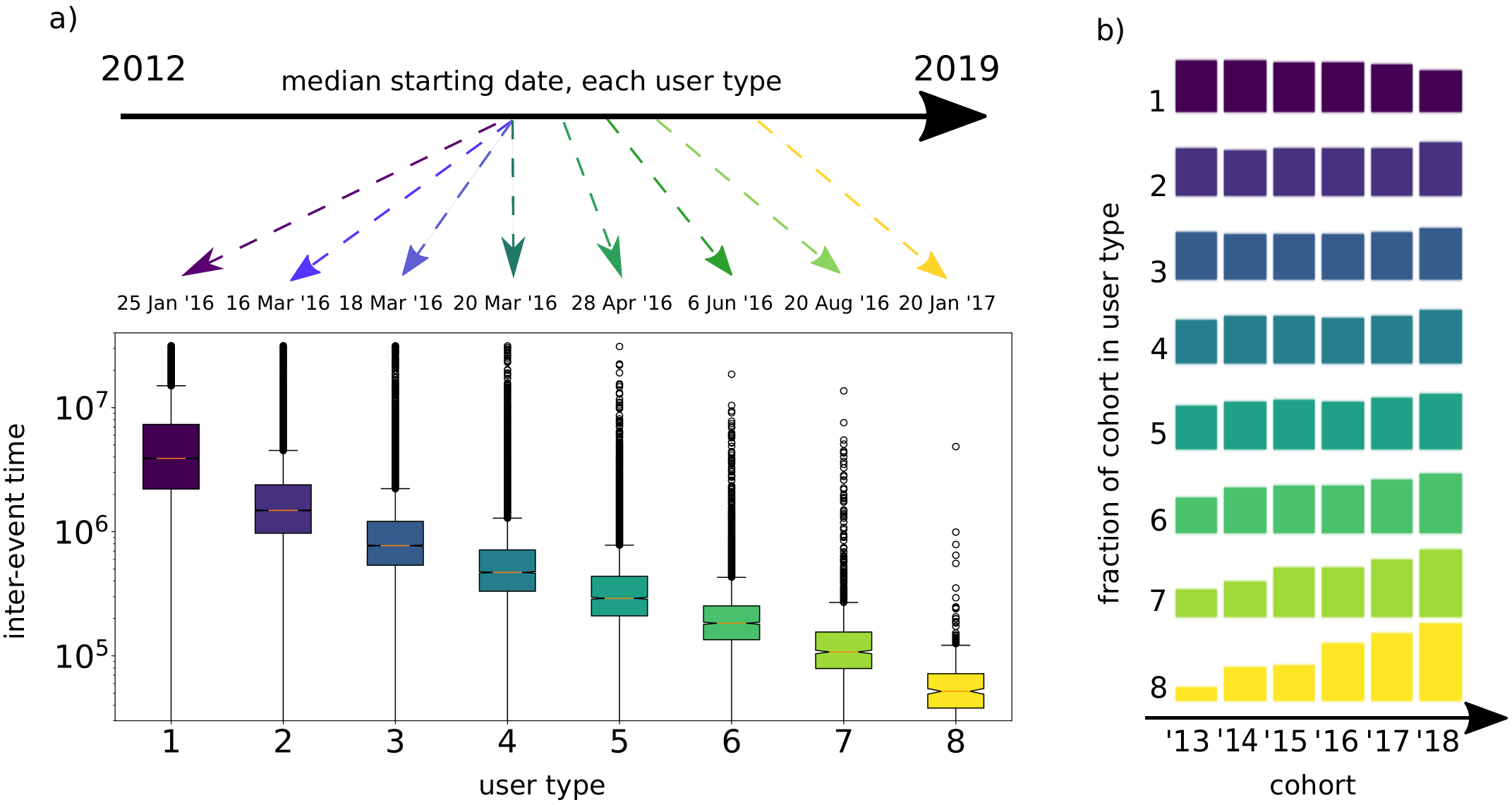}
\caption{Composition and inter-event time distributions in the first year of user types. a) User types are based on clustering along with the ratio of active days (at least one recorded tweet) versus total days on Twitter, labelled from 1 (least active) to 8 (most active). Top: Median starting dates (i.e., the first day of activity on Twitter $t_0$) for each user type. Bottom: Box plot representation of the individual inter-event time distribution for user types in seconds during their first year on Twitter. b) Composition of user types per cohort.}
\label{fig:2}
\end{figure*}

To better understand the different roles users might play in this process, we group users according to activity level. Our analysis indicates that the observed development on the cohort level stems from a changing composition of \emph{user types}, ranging from mostly passive spectators who tweet only occasionally to extremely active users (Fig.~\ref{fig:1}a). 

Activity was broadly distributed among users in each cohort (Fig.~S1a, c, e). 
To disentangle the heterogeneity of users, we compute the ratio of active days (fraction of days with at least one recorded tweet) to total days (days between the individual starting date $t_0$ and May 30, 2019) for each user (Fig.~\ref{fig:1}a). 
We identify eight user types via a simple, unsupervised $k$-means-clustering (using the Python package \emph{scikit-learn}\cite{scikit-learn}) and label them from 1 (least active) to 8 (most active). This method allows us to capture density variations in the data and thus set our bin-edges in a data-driven way.

Figure~\ref{fig:2}a shows the resulting separation of the user types regarding their activity. 
Note that, we here only consider the first year of activity (i.e., $t_0$) for each user to enable a meaningful comparison. Therefore, the upper bound of the inter-event time equals $31536000$ seconds (one year). Additionally, we emphasize that due to the 1\% random sample of tweets, inter-event times that we report cannot be easily interpreted as such but rather serve as a proxy for individual activity. 

Looking at user type and cohort, a trend becomes evident. 
Although we do not consider starting dates in the clustering we used to define user types, we find a striking one-to-one correspondence between activity and the median starting date for individuals belonging to each user type.
The more recent the median starting date, the more active the user type (see Fig.~\ref{fig:2}a, top). 
In particular, less active user types are only marginally separated in terms of their median starting dates while there is a pronounced shift towards more recent median starting dates for the more active user types. 
This points to a change in individuals' user types over time. 
To investigate this finding and understand how user types are distributed within each cohort, we determine the proportions of user types in each cohort. 

Figure~\ref{fig:2}b illustrates how the composition of user types changed over time. Perhaps the most striking observation is that the largest fractions of very active users can be found in later cohorts. 
The histograms show the proportions of user types in each cohort with respect to the absolute size of the cohort. 
Note that there are substantial differences between user types sizes  (user type 1: $61,052$ users, user type 8: $445$ users; for all numbers see SI Fig.~1d) and slight differences in cohort size. Therefore, absolute numbers are not well represented in Fig.~\ref{fig:2}b as we show relative size increase.

To exclude the possibility that these observations are mainly driven by automated accounts, beyond their relatively longevity in our sample, we compare highly active users to a randomly chosen set of users. To measure repetitive postings we use Shannon information \cite{chu2012detecting, cresci_bot2020} to quantify the complexity of the shared information. We find little difference in the distribution of complexity of tweets (on the word level) between very active user types and the random sample across all type (see Fig.~S8a,b). 

The evolving composition of user cohorts on Twitter highlights the fact that the most active user types grew, relative to the size of their group, more quickly than the less active user types did. Thus, the social acceleration observed on the collective level is likely to be driven by people who joined Twitter more recently and are using the platform differently compared to users who have been on Twitter longer.

\subsection* {Activity relates to network centrality}
\begin{figure}
\centering
\includegraphics[width=8.9 cm]{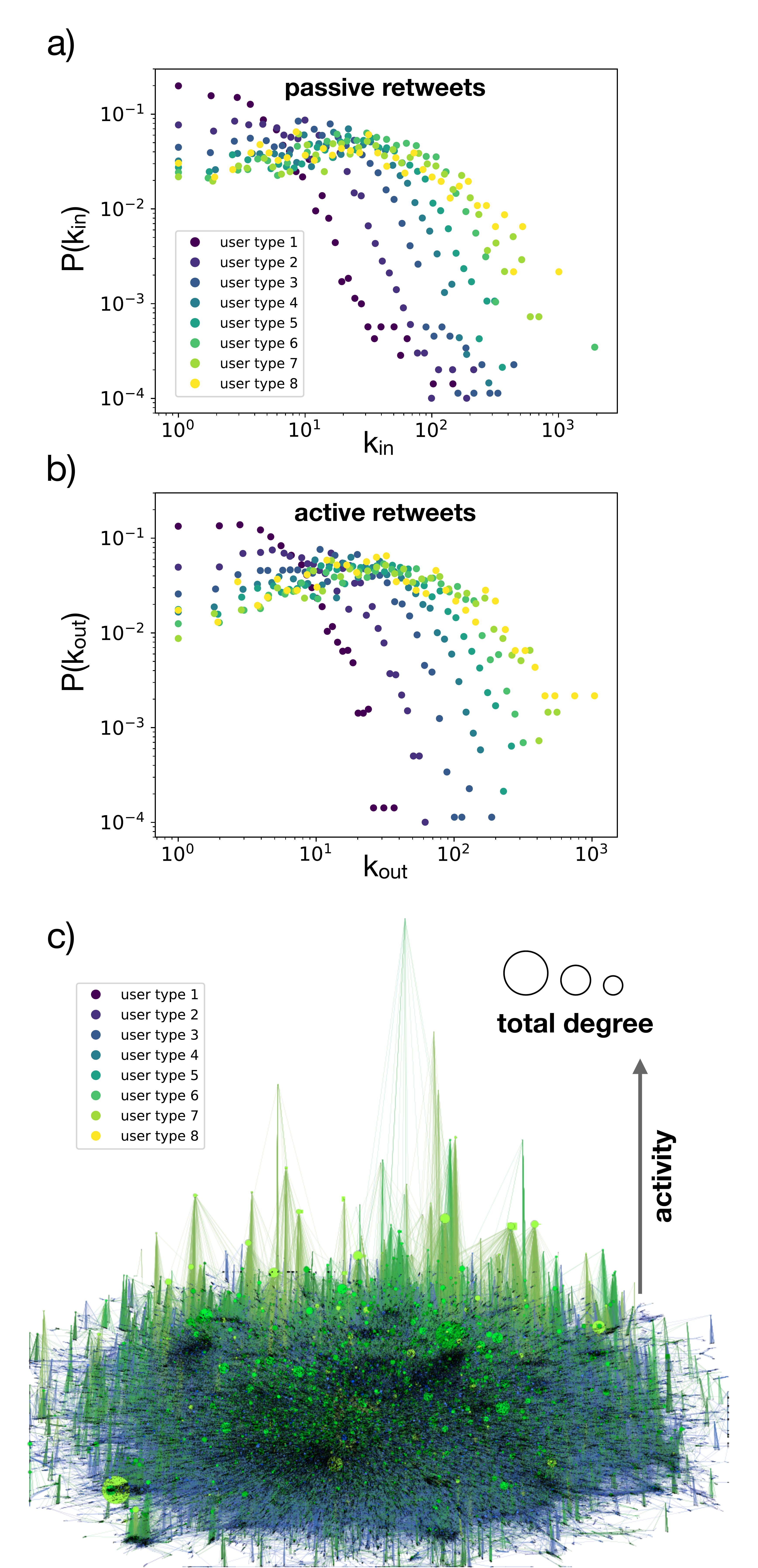}
\caption{Properties of retweet interactions. a) In-degree ($k_{in}$) distribution for each user type of the retweet interaction network, corresponding to passive retweets (i.e., being retweeted by someone else). b) Out-degree ($k_{out}$) distribution for each user type of the retweet interaction network, corresponding to active retweets (i.e., retweeting someone else). c) Illustration of the retweet network. Nodes represent users and are colored by user type. Diameter of the nodes varies according to their total degree; elevation along the z-axis shows activity (i.e., number of original tweets). Visualization via gephi \cite{Bastian2009}.}
\label{fig:3}
\end{figure}

To unpack the observed behaviour of the increasingly large fractions of highly active users further, next we focus on social interactions and how content changes for individuals. 
We find that highly active users are also well connected with others both actively (retweeting) and passively (being retweeted). 

To examine whether user types differed in respects other than their activity, we constructed a weighted network (aggregated over the full time range of $2012-2019$), consisting of both active and passive retweet interactions. For the construction of the network, we included retweet interactions outside of our random sample, by using each retweet from the full Twitter dataset whenever either the active or the passive user was from our sample. The network analyzed in Fig.~\ref{fig:3}, thereby consists of 648,880 unique users and a total of 5,963,284 interactions, resulting in 1,516,958 unique, weighted edges.

Figures~\ref{fig:3}a, b show the in- (Eq.~\ref{eq:indegree}) and out-degree (Eq.~\ref{eq:outdegree}) distribution of the eight user types. 
The increasing proportion of users with a high in- and out-degree indicates that more active user types are more central on both measures. 
Hence, more active users not only retweet more actively (which can be expected due to their on average higher overall activity) but also elicit more activity from other users. This is also reflected in the relatively high reciprocity of the network (0.743). Reciprocity is the ratio of retweets that were answered in the opposite direction at some point.
High reciprocity implies not only that activity increases over time (with more active users joining) but also that interactivity among users becomes more frequent. 
In other words, the observed trend towards higher activity does not appear to occur in isolation but may be connected to a collective effect of mutual social acceleration and denser interaction among Twitter users---for example, when an elevated level of interaction leads to more content appearing in users' feeds. 
Yet individual activity can also speed up social acceleration: by virtue of their high activity, users become more central, filling each other's feeds and thereby collectively contributing to the experienced social acceleration.

\subsection*{Content horizons}

Furthermore, and possibly driven by that growing overall activity, the amount of time any individual topic appears in people's tweets is shrinking. 
We call the amount of time that a topic tends to recur in a user's tweets, that person's \textit{content horizon}.
We operationalize this notion as the autocorrelation of hashtags an individual uses over time.

To set tweets in the context of an ongoing discussion, users employ hashtags, a combination of the ``\#'' symbol and keywords related to certain topics. Additionally, users can include URLs in their posts to link content to their activity. Did the way people interacted with content also change?

We find an increasing trend of sharing hashtags and URLs: 
Over our observation period, the numbers of hashtags per tweet and URLs per tweet almost doubled. 
While there is no clear difference among user types in terms of sharing URLs, more active user types tend to use more hashtags per tweet compared to less active user types (see Fig.~S3).
To understand the impact of the growing amount of interactions and content at an individual level, we measure the similarity of hashtags used over time and analyze an observable that we call the \textit{content horizon}, an adaptation of the broadly known concept of autocorrelation.

\begin{figure}
\centering
\includegraphics[width=8.9 cm]{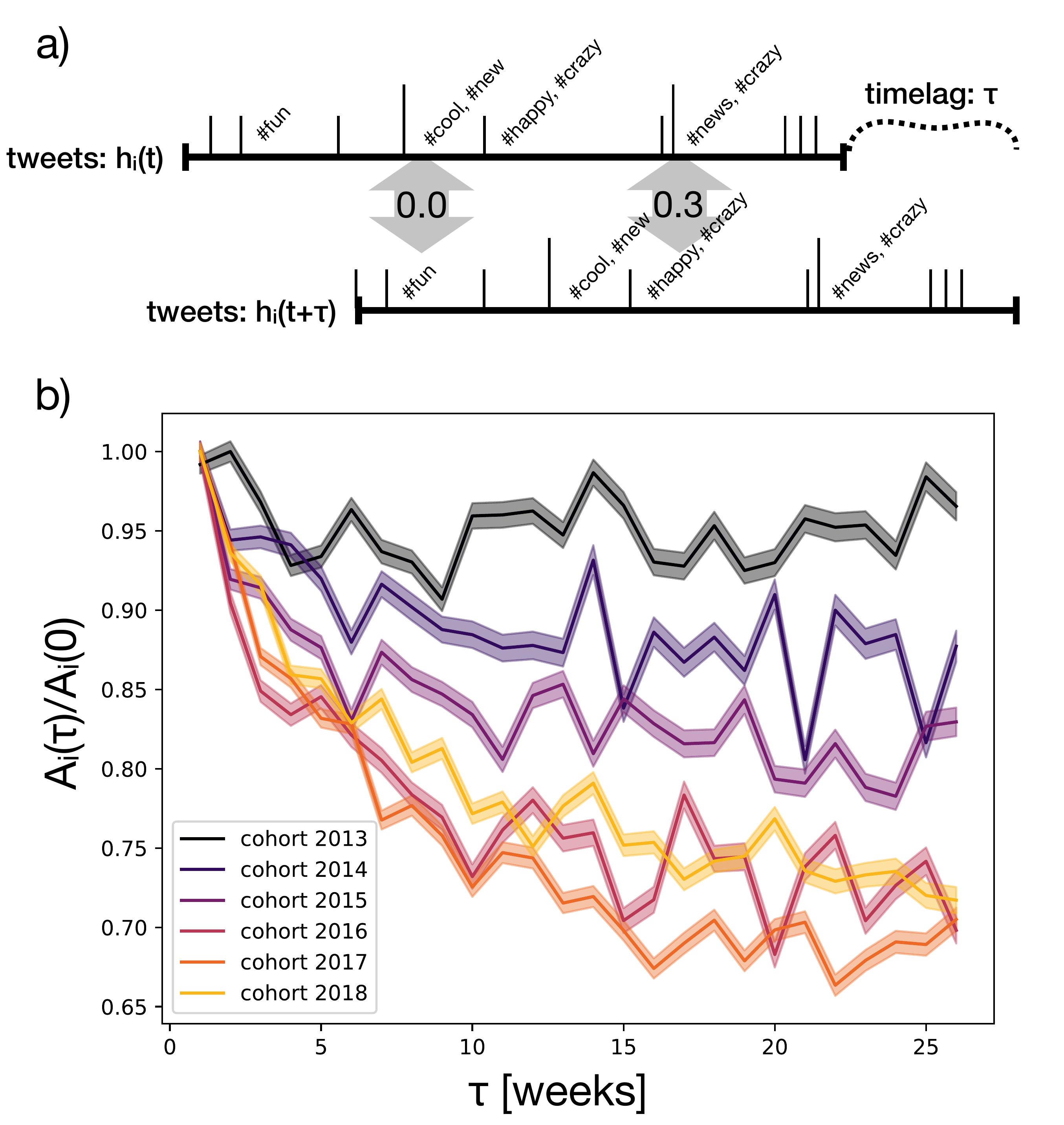}
\caption{Autocorrelation of hashtags. a) Illustration of the topical autocorrelation $A_{i}(\tau)$, via the relative overlap of hashtags between a shifted copy of one user's tweeting trajectory. b) The resulting autocorrelation as a function of an increasing time lag for the cohorts of Twitter users. Values are excluding the point without time lag $\tau=0$ and are relative to the maximum autocorrelation value for better comparison.}
\label{fig:4}
\end{figure}

To estimate the individual content horizons of Twitter users, we compute the individual autocorrelation function $A_{i,j}(\tau)$ of hashtags used in tweets (Eq.~\ref{eq:hashtagcorr}) as a proxy for characteristic length of time users focus on a specific topic before moving on (see top panel of Fig.~\ref{fig:4} for an illustration).

The decay of the autocorrelation is amplified in younger cohorts (Fig.~\ref{fig:4}). 
This change indicates that users in younger cohorts stopped tweeting about topics more quickly, potentially switching their focus to new topics. 
We call this development, which becomes stronger in successive cohorts of Twitter users, \emph{shrinking content horizons}. 
Because we observe the qualitatively same results for the different user types (see Fig.~S4), we assume that the shrinking content horizon is connected to rising activity on the part of individual users.

\section* {Discussion}

We found that the average activity per user on Twitter is increasing year over year. We observe this as a cohort effect and that this change can be explained by an effect of changing compositions of user types joining Twitter over time. Changes in individual trajectories of Twitter use are more difficult to measure due to the diversity of user type patterns on Twitter.
Because not only the sum of activities on Twitter is increasing \cite{lorenz2019accelerating}, but also the activity per user, the general development is likely to be due not only to a growing user base, but to actual differences in individual behaviour.
The change in the composition of user types is accompanied by growing connectivity among users through retweets, an increasing number of hashtags and URLs being shared, and shrinking content horizons (here operationalized as quickly decaying hashtag autocorrelations).

Even though we can not draw causal conclusions from our analysis, the combination of observations allows us to offer an explanation of the factors that are likely to contribute to an individual's experience of social acceleration on social media platforms: The growing fraction of highly active users in each successive cohort combined with the pronounced tendency for tweets to be retweeted within a single cohort leads to more interactions with other users, most likely resulting in more and more active users filling up their peers' feeds with content. 
This development is accompanied by a growing number of hashtags and URLs in tweets over time. 

Although the amount of content on Twitter is increasing, the amount of information on a Twitter feed that individual users can keep up with is finite---a simple consequence of the amount of information that fits on the screen in combination with users' attentional capacities \cite{hills2019dark}. 
Users, therefore, typically encounter an overabundance of content \cite{bawden2009dark}. 
The inability to keep up with everything that appears on their Twitter feed may explain higher turnover rates of topics for each individual. 
We quantified this development by measuring hashtag autocorrelation and confirmed the ever faster decay of individual content horizons. This points to a possible behavioural sequence of users in response to the growing information abundance: trading breadth for depth in their information behaviour \cite{carr2010shallows}
The shift towards spending less time discussing each topic, coupled with a simultaneous increase in the number of topics discussed, may be experienced by users as an aspect of social acceleration and is in line with previous results on the acceleration on the collective level \cite{lorenz2019accelerating}. 

However, our data and methods come with several limitations. As the data only includes active behaviour on Twitter it can not tell us much about the mechanisms at play that go beyond our analysis of interactions and content. For example, the data do not contain information about exposure (i.e., what people saw in their surrounding before they tweeted) on and off Twitter, or events outside of Twitter that drive activity exogenously \cite{burton2021reconsidering}. We aimed to exclude the possibility of a predominant presence of automated accounts via a complexity analysis of content from very active users, which does not seem to be overly repetitive. But while many other unobserved factors are important to better understand our observations, they do not trivialise our results; on the contrary, exterior factors are important to complement our findings and fill in the links between the dimensions that drive the broader development of social acceleration, only a small instance of which we could quantify here. 

For example, changes in the relative fraction of user types within each cohort over time could be an indication that the platform itself is changing: Platform design choices may be altering how Twitter users are motivated to interact \cite{brady2020mad, Lorenz-Spreen20, bak2021stewardship}. Other mechanisms that go beyond our measurements can include the increased professionalization and agenda-setting purposes of social media usage \cite{barbera2019leads} or an increasing migration of offline contacts to social media. Furthermore, future research could aim to connect the observed developments of acceleration, with attentional bottlenecks and the success of the spread of negative, emotional or hostile content on social media \cite{Rathjee2024292118, acerbi_2021, brady2017emotion, alvarez2015sentiment}.
To determine the drivers behind these potentially unintended developments in public discourse---to distinguish between algorithmic curation, amplified human tendencies, and societal developments---future research needs to examine such developments, also on other platforms. 
This research is only possible with access to data containing information about the platforms' sorting algorithms, design choices, exposure, as well as about observed behaviour \cite{pasquetto2020tackling}. 

Individuals are interacting with Twitter differently over time and influencing each other in the process. Our work provides an empirical starting point highlighting the need to quantify these complex, but important relationships further. Ultimately, the ability to quantify the directions in which the interplay of human behaviour, technological advancement and corporate interests drive online behaviour and discourse would help society to actively engage in shaping online discourse and identifying measures to promote a more deliberate online experience \cite{Lorenz-Spreen20, Kozyreva20}.

\section*{Materials and Methods}

\subsection*{Data set} 

We use a data set from Twitter's Decahose API. 
The data set features $10\%$ of the Twitter traffic between January 1, 2012 and December 6, 2016 and  $1\%$ between December 7, 2016 and June 6, 2019. 
To perform a continuous analysis across the abrupt change in temporal resolution, we employ only $10\%$ of the tweets in the first period, selected using random sampling (imitating the sampling mechanism of the Twitter API). Consequently, we consider $1\%$ of all tweets in the period between January 2012 and June 2019. A brief analysis of the full $10\%$ sample leading to comparable results is shown in the Supplementary Material, also reassuring that the different sampling methods do not dramatically affect our results \cite{pfeffer2018tampering}.

The metadata relates to each tweet consists of author ID, retweet ID, timestamp in seconds, and full tweet text. As the tweets and users are randomly sampled, there is no specific geographical or linguistic preference in the data set.

For our analysis, we set up three user samples, each containing the data from $200,000$ users. 
Users were selected randomly from all users who were active (at least one reported tweet) in March ($200,000$ users out of $28,883,037$), April ($200,000$ users out of $28,843,944$), and May 2019 ($200,000$ users out of $29,353,930$). 
The user samples are almost distinct as the pairwise overlap between the samples is less than 0.5\%. Only two users appear in all three data sets.  
This user selection allows us to track the individual user activity over the whole time period (Jan 2012 - Jun 2019) by considering all tweets from each selected user in the $1\%$ subset and it assures that there is no drop-out of users in the studied period. Of course, this limits our analysis to those users who have remained on the platform until Spring 2019. There are, however, good reasons to believe that this selection does not strongly affect our conclusion: users who dropped out may have behaved systematically differently, but as these users are more likely to be from early cohorts, and the assumption that they were less active on Twitter before they dropped out would imply that differences in activity are possibly higher than reported in our work.

In the main text, we show the results for the sample of users based on tweets recorded in April 2019. 
We establish that our results are stable for other user samples obtained in a distinct period in the Supplementary Material, by showing results from analyzing the activity of users in the other two samples. The choice of using the sample from April 2019 for the main manuscript is arbitrary and aside from slightly different user type compositions, all results from all samples agree quantitatively and qualitatively (see also the Supplementary Material).

\subsection*{Network analysis}
We encode the topological information of a network with $N$ nodes (here representing users) in the weighted adjacency $\boldsymbol{W}$. 
Matrix elements $w_{ij}$ of $\boldsymbol{W}$ indicate how often user $i$ has retweeted tweets from user $j$ \cite{Strogatz2001, Newman2003}.

Hence, the in-degree is defined by \cite{Barabasi2016}
\begin{equation}\label{eq:indegree}
    k^i_{in} = \sum_{i=1}^N w_{ji}.
\end{equation}

The out-degree is defined by \cite{Barabasi2016}
\begin{equation}\label{eq:outdegree}
    k^i_{out} = \sum_{i=1}^N w_{ij}.
\end{equation}

Active retweets are therefore represented by the out-degree (Eq.~\ref{eq:outdegree}), and passive retweets (e.g., if user $A$'s tweet was retweeted) are represented by the in-degree (Eq.~\ref{eq:indegree}).

\subsection*{Hashtag autocorrelation}
We pursue a nonstandard approach and employ Jaccard similarity to enable comparing the categorical hashtag data. Specifically, we define the lagged correlation between two tweets at $t$ and $t+\tau$ of an individual user $i$ as

\begin{equation} \label{eq:hashtagcorr}
 A_{i}(\tau) = \frac{1}{N_{\text{match}}} \sum_{t=0}^{T-\tau} J(h_i(t), h_i(t+\tau)).   
\end{equation}
Here, $T-\tau$ determines the number of weeks that the user has been active on Twitter and $J(A,B)= \frac{A\cap B}{A\cup B}$ represents the Jaccard similarity.

Zero correlations can be caused by different settings such as no activity, no used hashtags, and no common hashtags. To avoid an activity bias in the correlation measure, we exclude weeks in the computations during which one of the users $i$ and $j$ had been inactive and normalize only by the number of non-zero entries (i.e., $N_{\text{match}}$).

\noindent \textbf{Supplementary Material:} accompanies this paper at {\small {\tt http://www.scienceadvances.org/}}.

\bibliography{ai_online.bib}
\bibliographystyle{ScienceAdvances}

\noindent \textbf{Acknowledgements:} 
We thank Deborah Ain for editing the manuscript and the research group at the Center for Adaptive Rationality and Thomas Peron for helpful comments and discussions.
\noindent \textbf{Funding:} FW acknowledges financial support from the International Research Training Group 1740/TRP 2011/50151-0 (funded by the German Research Federation and the S\~ao Paulo State Foundation for the Promotion of Research) and the German Ministry of Education and Research (BMBF) via the project ClimXtreme (grant no. 01LP1902J). PLS acknowledges financial support from the Volkswagen Foundation (grant ``Reclaiming individual autonomy and democratic discourse online:  How to rebalance human and algorithmic decision making''. SL acknowledges support from the HOPE project (Carlsberg Foundation) and the Nation Scale Social Networks Project (Villum Foundation). \\
\noindent \textbf{Author Contributions:} All authors designed the study. FW and PLS evaluated the data. All authors analyzed the results and wrote the manuscript.  FW and PLS contributed equally to this work.
\noindent \textbf{Competing Interests:} The authors declare that they have no competing financial interests.\\

\clearpage

\end{document}



\baselineskip24pt


\maketitle



In the main manuscript, we have based our analysis on the tweeting behavior of $200,000$ users in who we randomly selected among all users who at least posted a single tweet on Twitter in April 2019. In this supporting information we do not only show figures regarding specific findings of this user group but also illustrate the robustness of the results by showing the raw data and comparing the outcomes with two other independently obtained data sets of the same size. These two other data sets are based on user activity (users randomly sampled from all users who tweeted) in March 2019 and May 2019, respectively. In the following we refer to the sample from March 2019 as sample 1, to the sample from April 2019 (used for the main analysis) as sample 2 and to the sample from May 2019 as sample 3.

\section*{Inter-event times of user cohorts and user types}

\subsection*{Cohorts}
In the main manuscript, we showed the rising activity of user cohorts in Figure 1. We based this measure on the cohort-mean number of tweets per week. A measure that is directly related to the user activity is the inter-event time. In Fig.~S\ref{SIfig:1}a,c,e we show the inter-event times for all cohorts of the data sets based on user activity in March 2019 (a), April 2019 (c) and May 2019 (e). All three panels depict the increasing share of short (sub-daily) inter-event times and, thus, the rising activity. For all data sets there is an observable change towards shorter inter-event times for cohorts with users having joined Twitter more recently. This results do not only further confirm the findings stated in the main manuscript but underline its robustness as we find comparable results for the other two data sets.

\subsection*{User types}

\begin{figure}
\centering
\includegraphics[width=\textwidth]{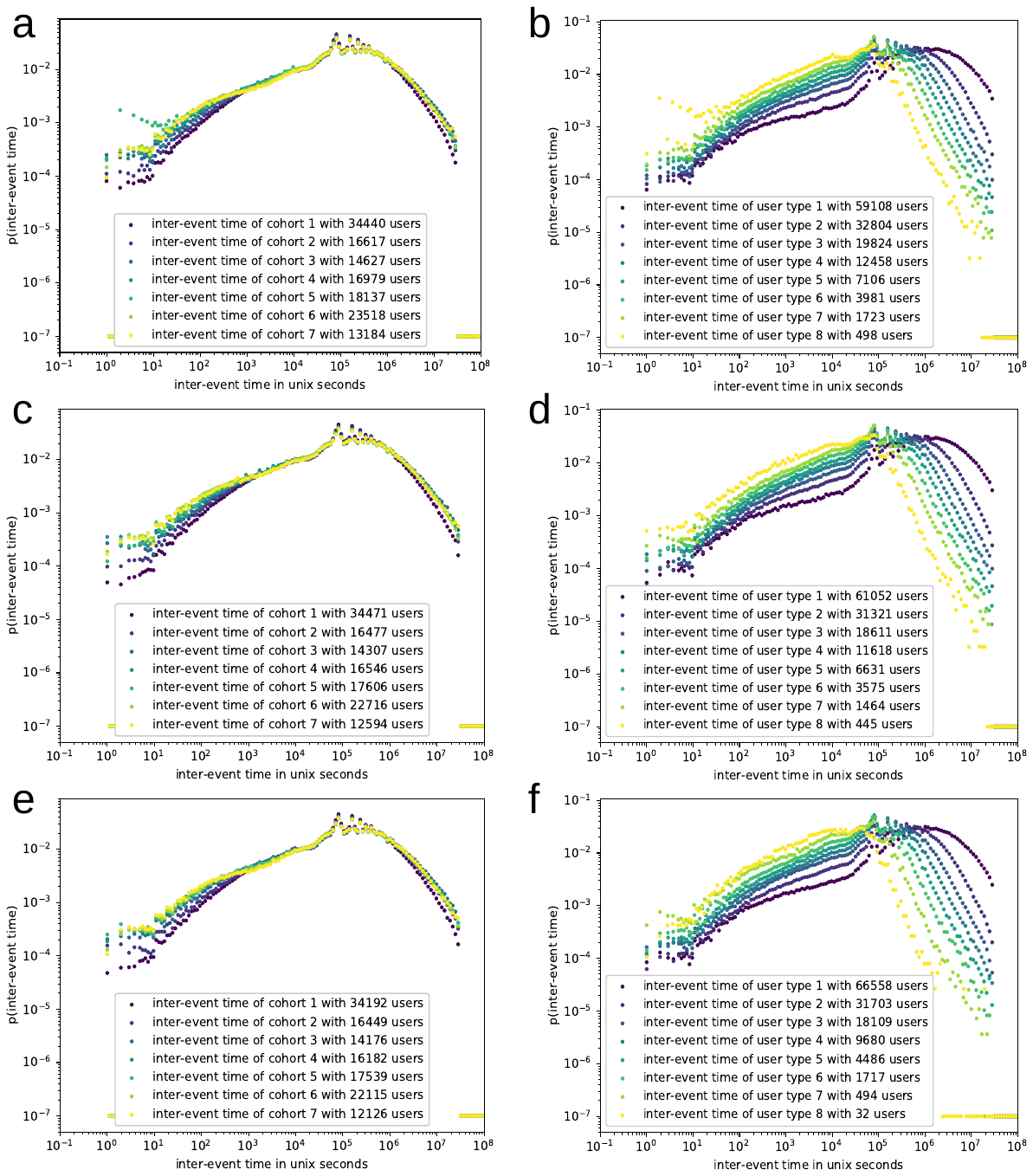}
\caption{Inter-event times for cohorts (a,c,e) and user-types (b,d,f) for all three samples (top to bottom).}
\label{SIfig:1}
\end{figure}

In the main manuscript we spilt the users into user types of distinct activity by clustering the users by means of the fraction of active days. In Figure 2a of the main manuscript we confirmed, that the distributions of the inter-event times of the user types differ regarding their quantiles. In Fig.~S\ref{SIfig:1}b,d,f we show the inter-event times of the distinct user types for all three data sets (panel b shows the distribution from the data set studied in the main manuscript). The results confirm the robustness of the results as all three data sets show the same characteristic differences between the user types and agree regarding the overall distribution.

\newpage
\section*{Median starting date and url/hashtag use}
\begin{figure}
\centering
\includegraphics[width=0.9\textwidth]{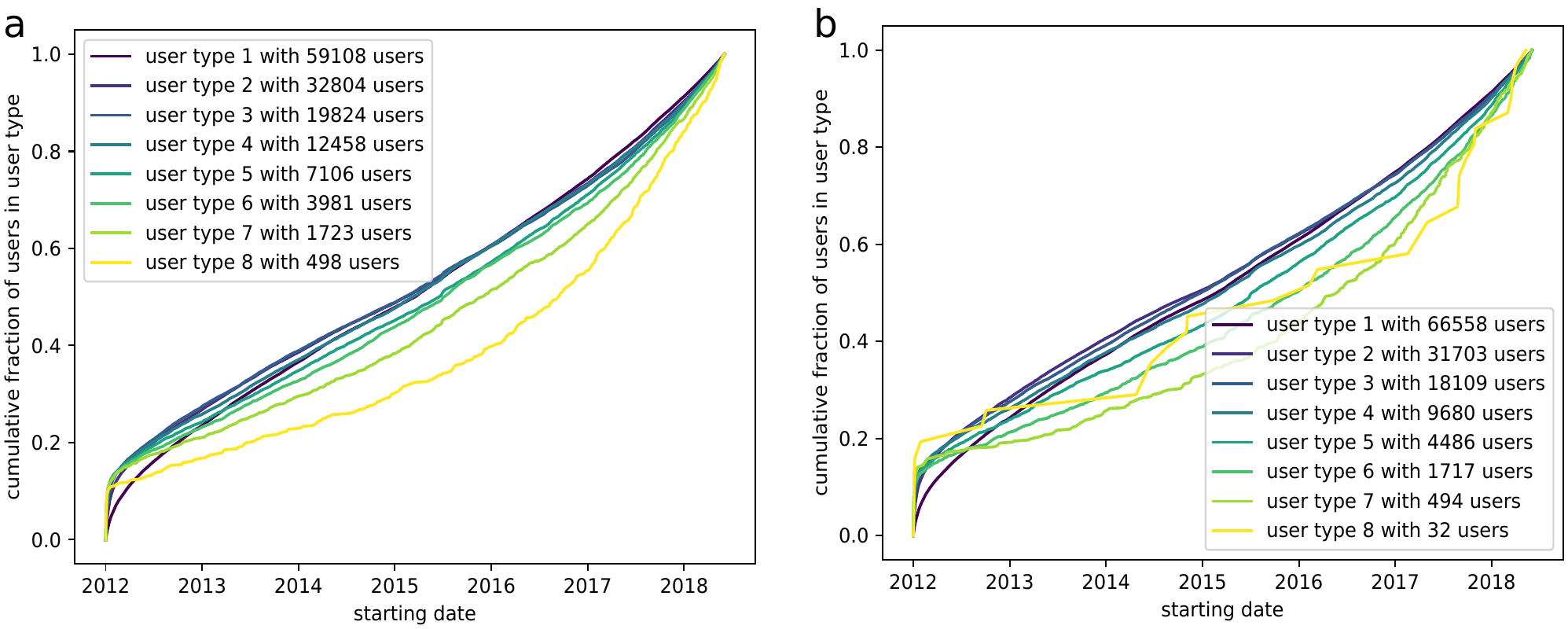}
\caption{Cumulative age distribution of users in the different user-types in sample 1 (a) and sample (3). Note that we here include users from year 2012 (which we have not included in the calculations for Figure 2 in the main manuscript. Panels (c)-(f) show the average hashtags (c,d) and URLs (e,f) for sample 1 and sample 3.}
\label{SIfig:2}
\end{figure}

As an additional feature to the systematic offset between the user types, we studied the median starting date of the user types in Figure 2 of the main manuscript. To underline the robustness of our identified correspondence between user type activity and median starting date we show the cumulative starting date distribution for the two other data samples in Fig.~S\ref{SIfig:2}a,b. For both samples we confirm that the user types exhibit a clear correlation between median starting date and activity as well as the increasing fraction of recent users in the more active user types.

\begin{figure}
\centering
\includegraphics[width=0.8\textwidth]{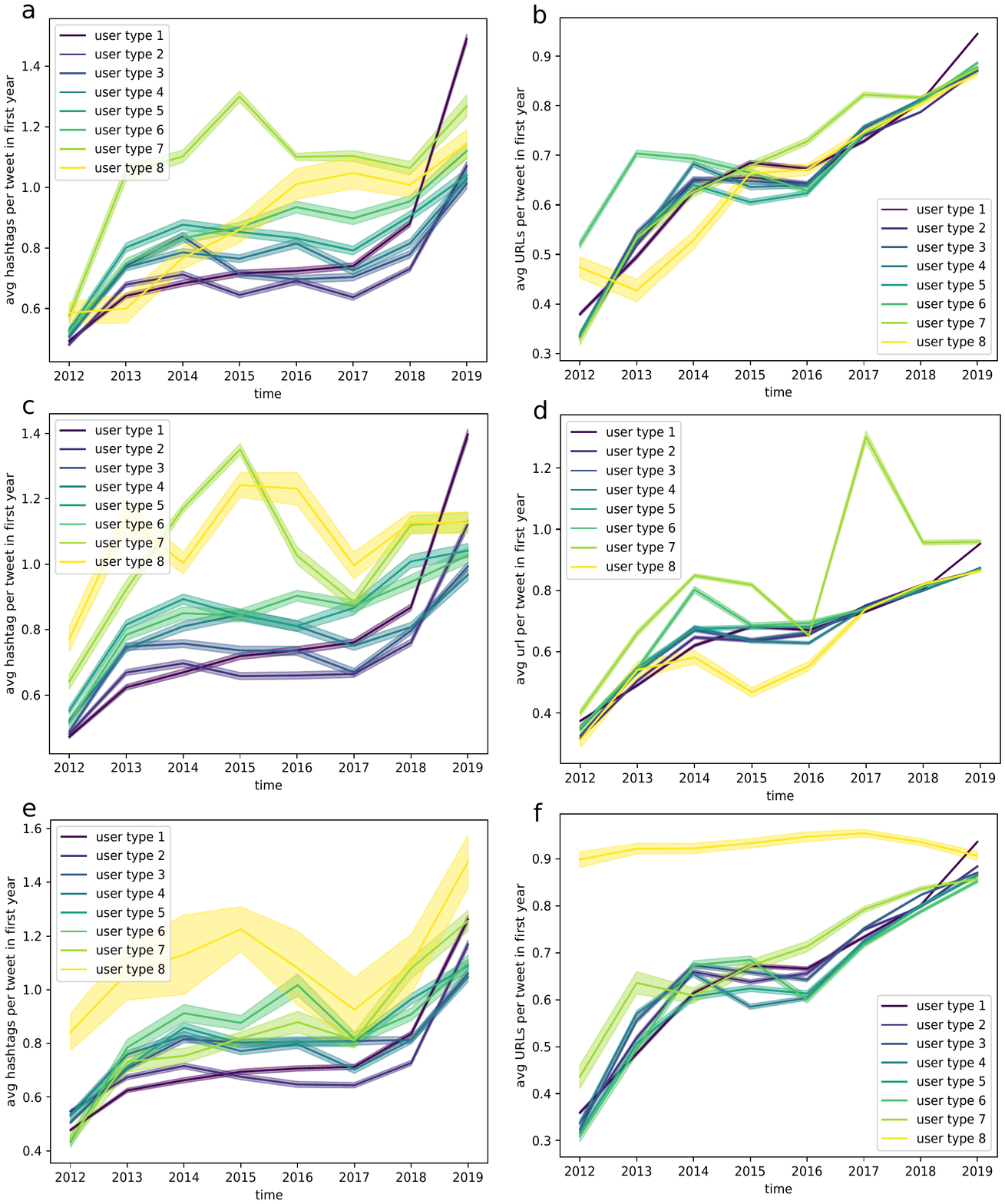}
\caption{Average number of hashtags (a,c,e) and URLs (b,d,f) per user-type in sample 1 to 3 (top to bottom). Shaded areas indicate standard deviation.}
\label{SIfig:3}
\end{figure}

The general increase of user activity goes in hand with a simultaneous increase of hashtag and URL usage. For the all three samples we show the average numbers of URLs and hashtags per tweet in Fig.~S\ref{SIfig:3}.

\newpage
\section*{Content horizons by user types}

\begin{figure}
\centering
\includegraphics[width=0.7\textwidth]{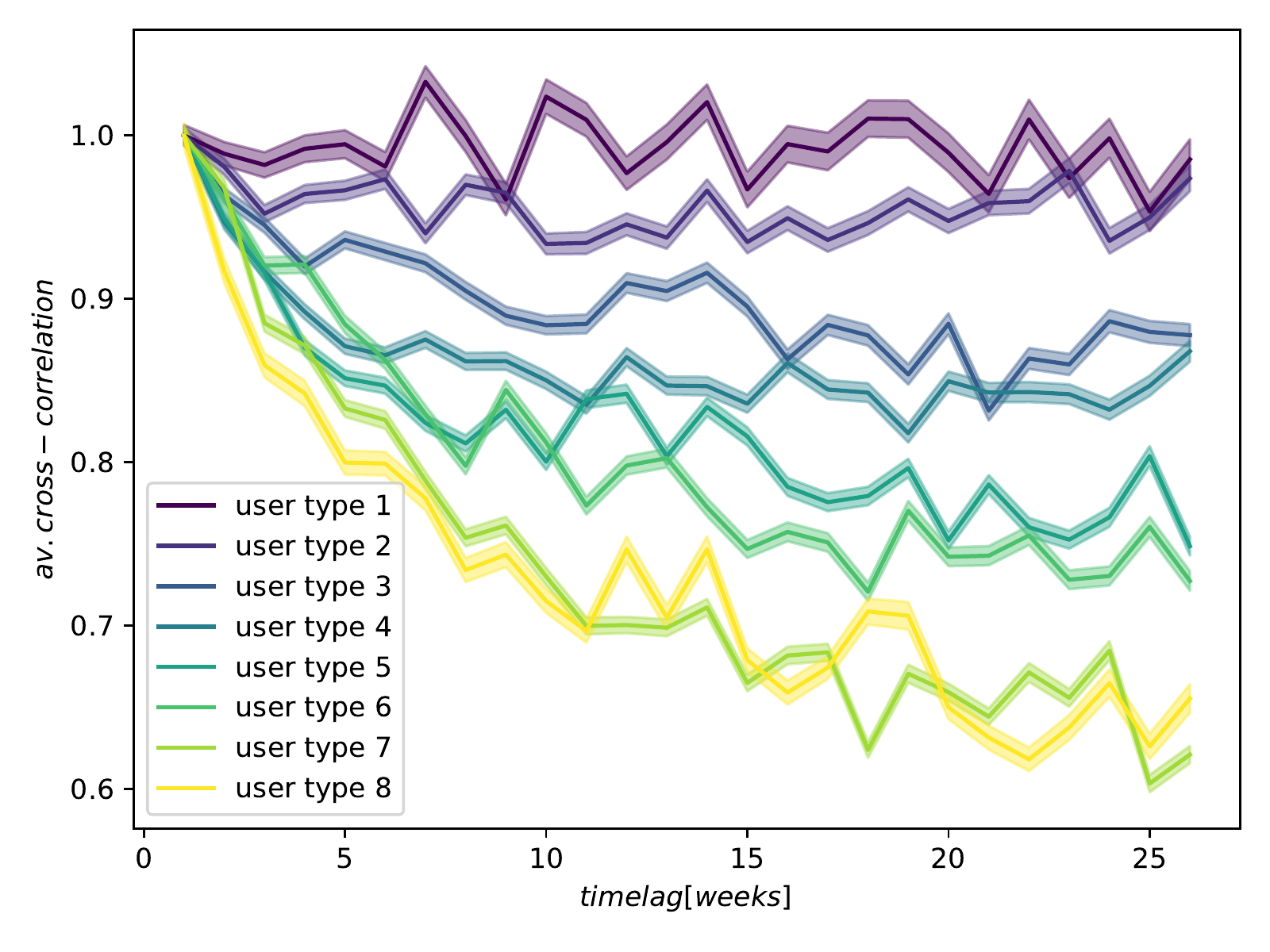}
\caption{Content horizon by user types for user types in sample 2.}
\label{SIfig:4}
\end{figure}

As mentioned in the main manuscript, we observe a shrinking content horizon not only for the cohorts but also for the user types (with a even stronger decaying autocorrelation function). To illustrate that, we show the content horizon for the user types of sample 2 in Fig.~S\ref{SIfig:4}.

\newpage

\section*{Analysis of full 10\% data set between 2012 and 2016}
\begin{figure}
\centering
\includegraphics[width=0.7\textwidth]{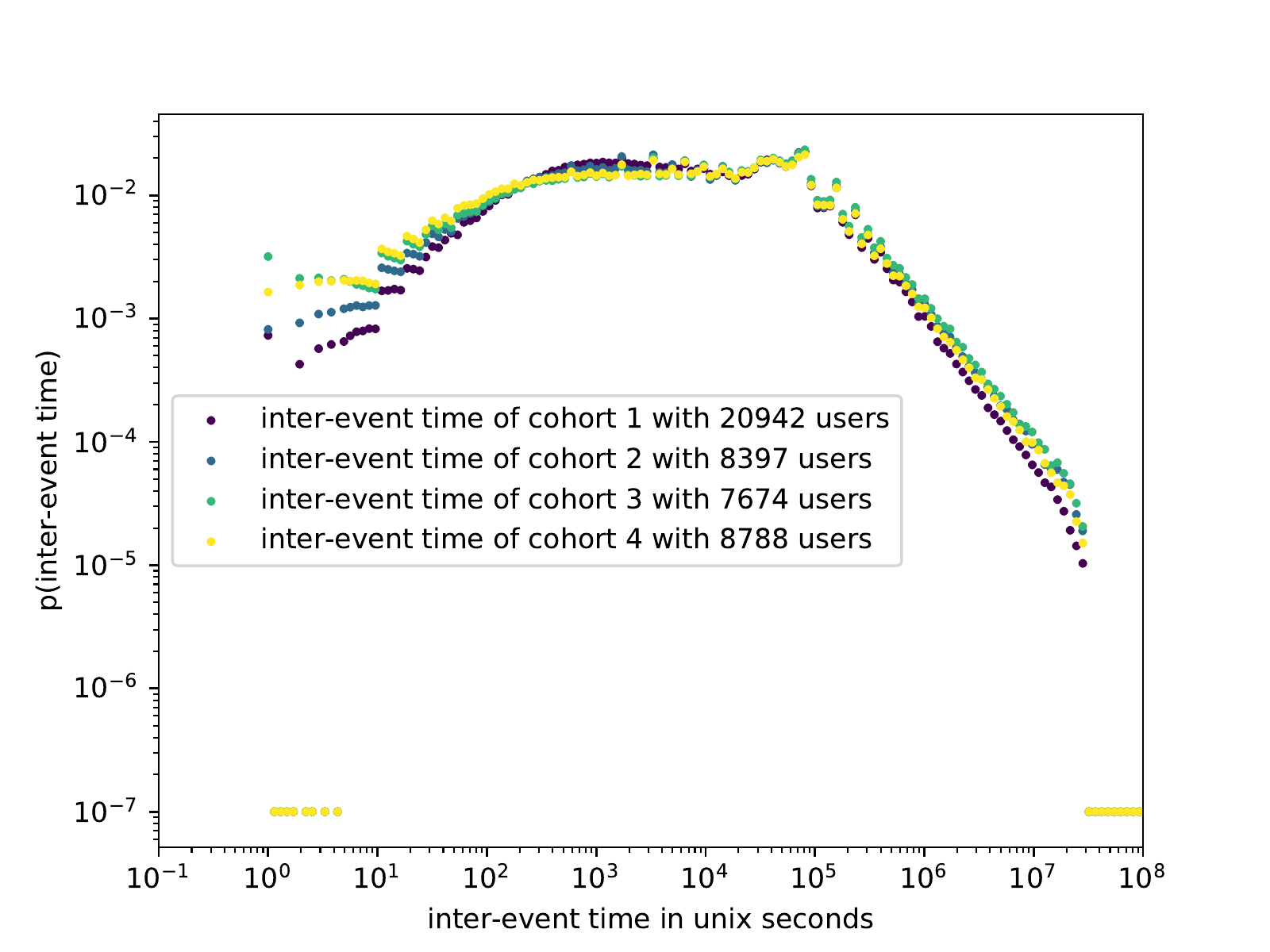}
\caption{Inter-event time distribution using the full 10\% data set for users in sample 2.}
\label{SIfig:5}
\end{figure}

\begin{figure}
\centering
\includegraphics[width=\textwidth]{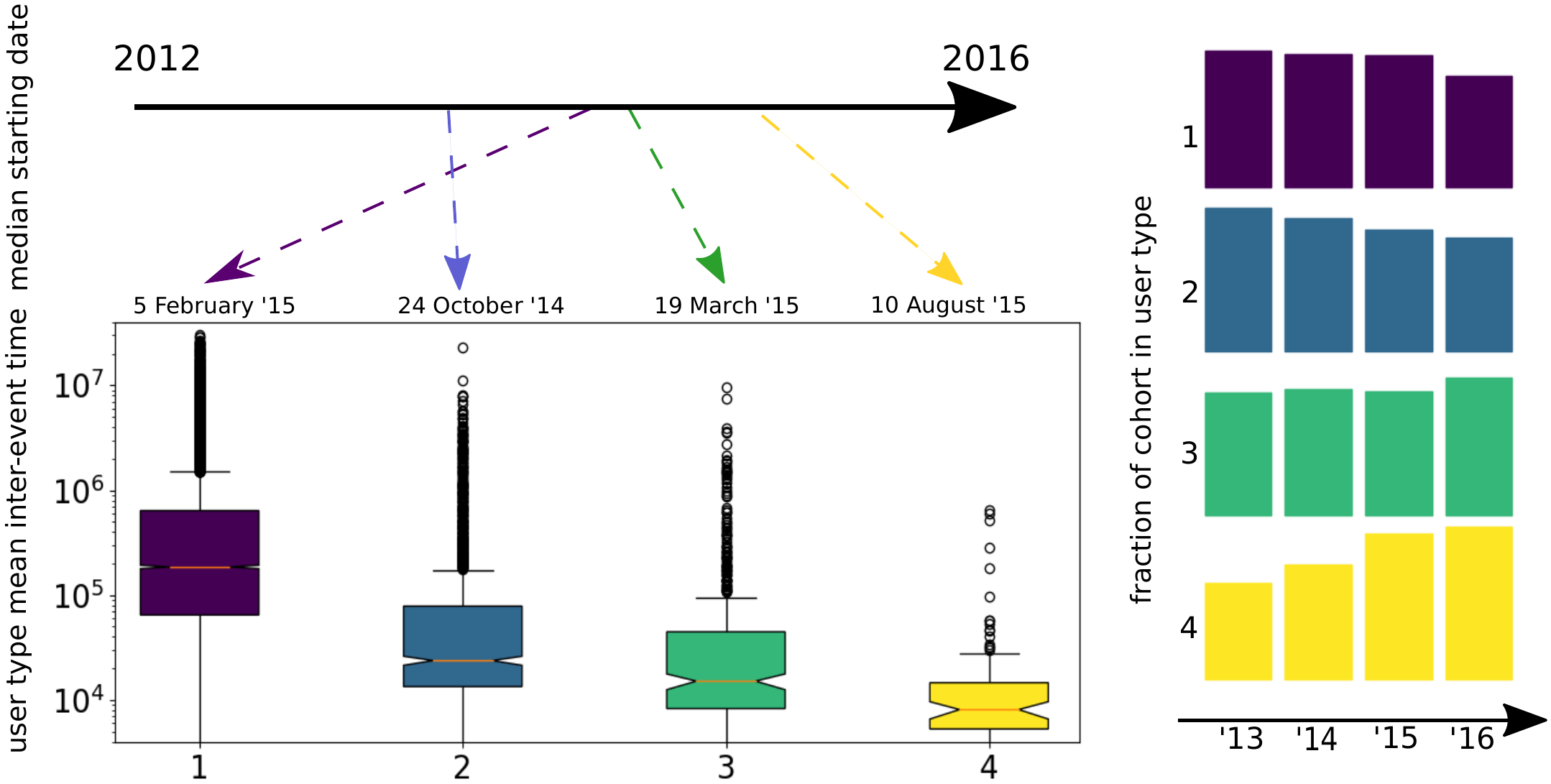}
\caption{Composition and inter-event time of user types. a) User-mean inter-event time during the first year on Twitter and median starting date for each user type. b) Cohort compositions of user types over time. Both is shown using the full 10\% data set for users in sample 2.}
\label{SIfig:6}
\end{figure}

Additionally, we want to shed light on the downsampling procedure which enabled us to compare and contrast Twitter user behavior over the full study period. We are aware that reducing the density of the tweets can substantially affect our results. For assuring that the implications our our analysis would have been the same utilizing a data set with a larger fraction of tweets, we utilize the full 10\% data set between 2012 and 2016. For this data set, we conducted the same analysis as for the whole data set. In Fig.~S\ref{SIfig:5}, we show the inter-event time distribution of the four cohorts of this data set. Although the differences are not as striking as for the whole data set (which is caused by the shorter period covered), the same trend of an increasing share of shorter inter-event times can be observed. 
In addition, we obtained 4 user types of differing activity employing the same procedure as for the three other data sets (k-means clustering based on ratio of active days). Our results indicate the described observation: The relative fraction of the more recent users increases in when comparing less active and more active user types (see Fig.~S\ref{SIfig:6}.

\newpage

\section*{Stability test with random users}
\begin{figure}
\centering
\includegraphics[width=0.7\textwidth]{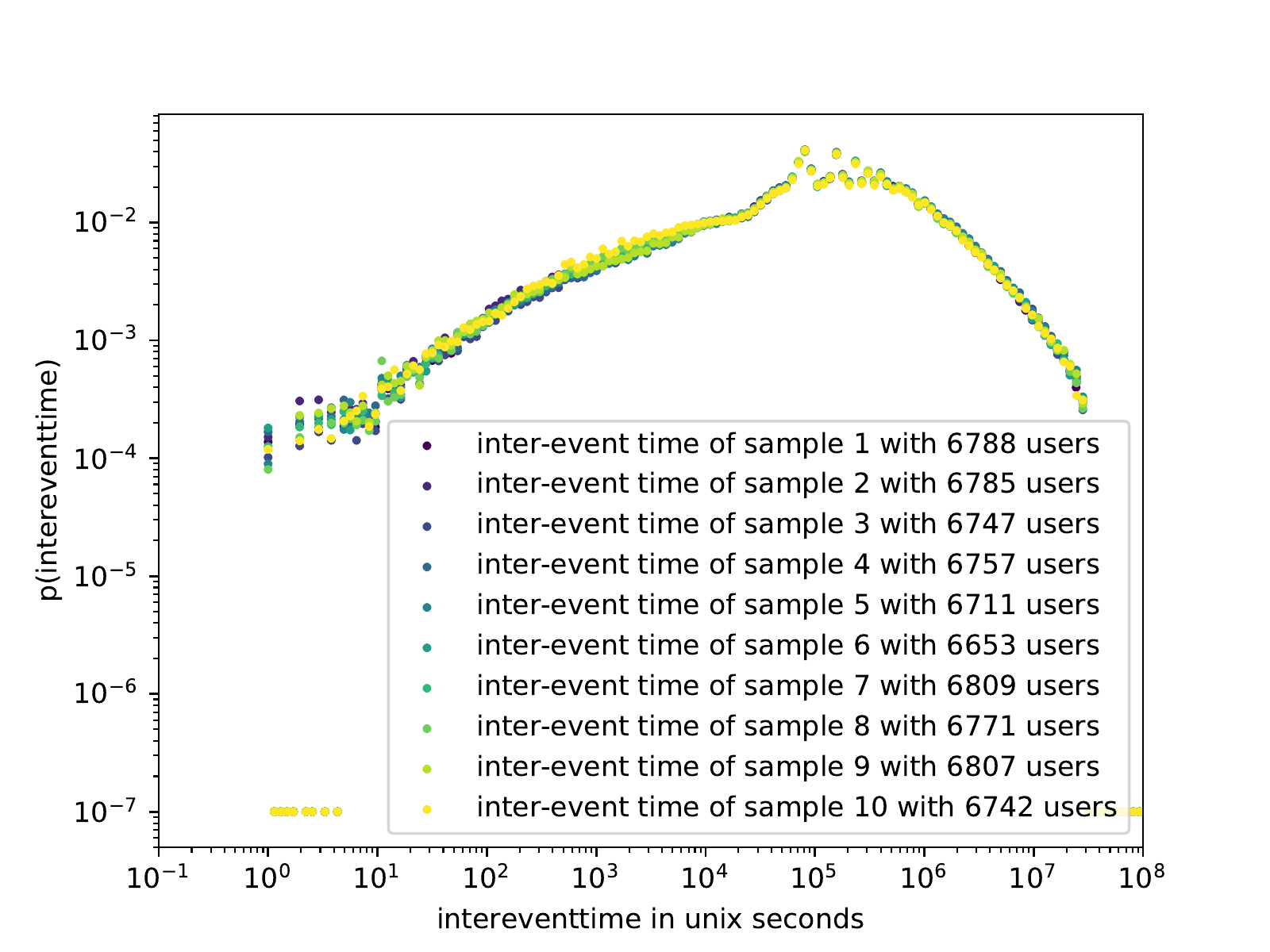}
\caption{Inter-event time distribution of ten random user samples drawn from sample 2.}
\label{SIfig:7}
\end{figure}

To assure that the differences observed between the cohorts and user-types has not been caused by coincidence, we selected random samples of $10000$ users from sample 2 and computed the inter-event times of each user to analyze the sample-wise inter-event time distribution. Note that we only include the inter-event times of all users which have tweeted for a full consecutive year and only take first-year inter-event times into account. Therefore, we exclude $ \sim \frac{1}{3}$ of the random samples. As indicated in Fig.~S\ref{SIfig:7} all distributions collapse on one common and there is no systematic difference between the sub-samples. This is not only a confirmation that observing differences as shown in the previous figures is very unlikely. We also show that already a sample size of $\sim 7000$ users allows for obtaining the full typical distribution of inter-event times. We consider that as an additional validation of the significance of our results.

\newpage
\section*{Bot activity}
In our analysis, we showed that the general increase of activity on Twitter is related to a growing share of the more active user types. On suspect might be that these highly active users are mostly bots tweeting nonsense at a high rate and, thus, contribute to the measured increase of activity.  For comparing the content of highly active users with normal users, we randomly sampled $1000$ users from the most active user types of sample 2 ($200$ of the most active user type, $400$ of the second and third most active user types) and $1000$ random users of sample 2. As we are only analyzing $1\%$ of all tweets standard approaches like measuring the (unnatural) temporal distance between tweets does not easily work. As a first naive approach, we analyzed content repetition. Repeating content extensively can be associated with bot activity. $737$ of $1000$ users of the active user types have never repeated a post and 950 have more than $90\%$ original posts. In the control group of 1000 random users we found that $847$ users have never repeated a post and 985 users have more than $90\%$ original posts. As the great majority of highly active users posts original content on Twitter, we confirm that there are no direct implications of bots significantly biasing the results. 

To complement this basic analysis with a more sophisticated approach, we decided to measure the information of words that each user has included in their tweets. Therefore, we first assembled all tweets of each individual user. To measure the complexity of the words this user has posted over the whole time this user has been active on Twitter, we have then quantified the information by considering each word as a single entry of a dictionary.  Assuming that we have a dictionary of user $i$, $ \mathcal{D}_i = \{\text{word}_1,\text{word}_2,...,\text{word}_n\}$, we quantify the information of this dictionary as

\begin{equation*}
    I_{total} = \sum_{i=1}^n m_i*I(\text{word}_i)
\end{equation*}
with $I(\text{word}_i) = - \ln{p_{\text{word}_i}} $ and $m = 100* p_{\text{word}_i}$. $p_{\text{word}_i}$ denotes the ratio of how often a particular word has been used to the total number of words that this user has tweeted. 

\begin{figure}
\centering
\includegraphics[width=\textwidth]{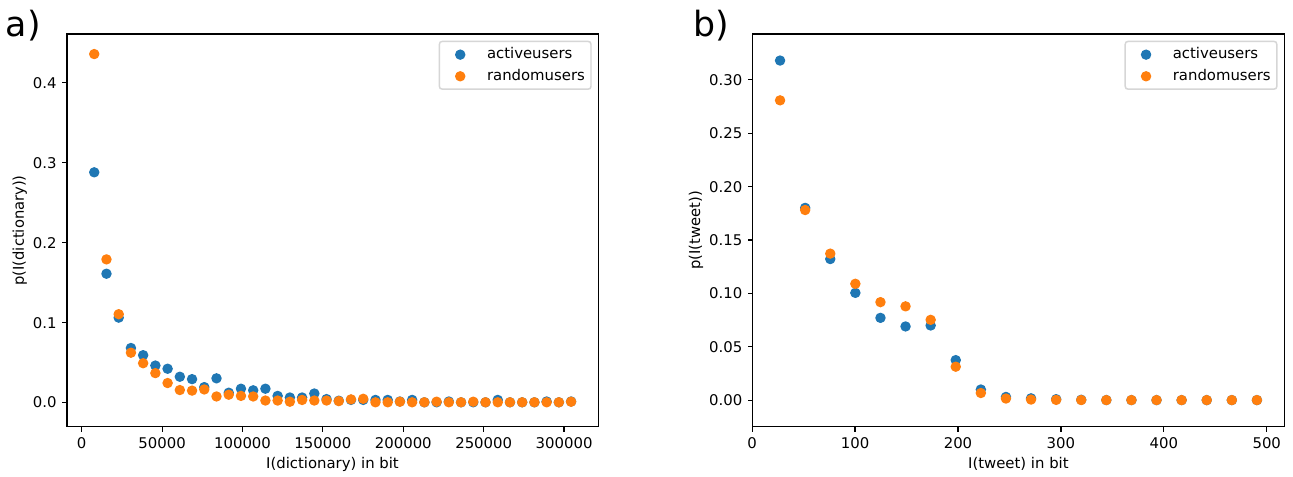}
\caption{Inter-event time distribution of ten random user samples drawn from sample 2.}
\label{SIfig:8}
\end{figure}

 Figure S\ref{SIfig:8}a shows the distribution of total information per dictionary for the active and the random user sample. Apparently, the more active users have more rich dictionaries. This is simply due to the fact that a dictionary with more words leads to a higher total information. But already from this, we can conclude that the majority of the highly active users tends to use more different words than the average user which is a strong indication of natural behavior which we do not expect from bots.

To further quantify how the users in the two selected groups use their own dictionary to formulate more or less complex tweets, we computed the information of each tweet of each user given the dictionary of this user. The information contained in tweet $T$ with $l$ words is computed as

\begin{equation*}
     I_{T} = \sum_{i=1}^l I(\text{word}_i).
\end{equation*}

Figure S\ref{SIfig:8}b shows the distribution of the information per tweet for the different user groups. At the first sight, both distributions share similar features. There is a monotonous decrease towards high information per tweet and a small plateau for medium information. The only minor difference is the slightly larger share of tweets containing almost no information of the highly active users. We here want to emphasize that this does not necessarily mean that a user tweets nonsense or always the same; it rather refers to users having a large dictionary and only using the often occurring words for a majority of posts. This is exactly the behavior we would expect from highly active but human users. 

Summarizing, the analysis of the semantics of the tweets revealed that there is no directly hint towards the suspicion that the highly active users are mostly bots. Of course, we cannot conclude that none of the users with a high activity is a bot (indeed we assume that there are bots active and included in our samples) but we believe that given the high similarity of the semantic information in tweets by random users and highly active users our results are not predominantly driven by bot behavior.

\clearpage